\documentstyle[12pt]{article}
\begin{document}

\begin{titlepage}
\vspace*{1cm}
\begin{center}
{\Large \bf Theory of Photoluminescence from a Magnetic Field
Induced Two-dimensional Quantum Wigner Crystal}

\vspace{0.5cm}
{\bf D.Z. Liu}\\
{\it Center for Superconductivity Research\\
Department of Physics\\
University of Maryland\\
College Park, Maryland 20742}

\vspace{0.5cm}
{\bf H.A. Fertig }\\
{\it Department of Physics and Astronomy\\
University of Kentucky\\
Lexington, Kentucky 40506-0055}

\vspace{0.5cm}
{\bf S. Das Sarma}\\
{\it Department of Physics\\
University of Maryland\\
College Park, Maryland 20742}

\vspace{1cm}

\end{center}

\begin{abstract}

We develop a theory of photoluminescence
using a time-dependent Hartree-Fock approximation that is
appropriate for the two-dimensional Wigner crystal
in a strong magnetic field.  The cases of localized
and itinerant holes are both studied.  It is found that the photoluminescence
spectrum is a weighted measure of the single particle density of states
of the electron system, which for an undisturbed electron lattice has the
intricate structure of the Hofstadter butterfly.
It is shown that for the case of a localized hole,
a strong interaction of the hole with the electron lattice tends to wipe out
this structure.
In such cases, a single final state is strongly
favored in the recombination process, producing a single line in the spectrum.
For the case of an itinerant hole, which could be generated in a wide
quantum well system,
we find that electron-hole interactions
do not significantly alter the density of states of the Wigner crystal,
opening the possibility of observing the Hofstadter gap spectrum
in the electron density of states directly.  At experimentally
relevant filling fractions, these gaps are found to be extremely
small, due to exchange effects.  However, it is found that
the hole, which interacts with the periodic potential of the
electron crystal, has a Hofstadter spectrum with much larger
gaps.  It is shown that a finite temperature experiment would
allow direct probing of this gap structure through photoluminescence.

\end{abstract}
\end{titlepage}

\begin{center}
{\bf I. INTRODUCTION}
\end{center}

   The search for a magnetic field induced Wigner crystal(WC) in
two-dimensional electronic systems has been a subject of long-standing
interest.
It was first pointed out by Wigner\cite{wigner}
that lowering the density of a
quantum electron system would lead to crystallization, since quantum
fluctuation effects diminish more rapidly than Coulomb correlation.
The relevent comparison is the Coulomb interaction energy
$V_{c}=e^{2}/\epsilon a$ to the zero-point energy
$K=\hbar^{2}/m^{*}a^{2}$, where $a=(\pi n)^{-1/2}$ is the mean
inter-electron distance and $\epsilon $ is the dielectric constant of
the host material. Defining the ratio $r_{s}=V_{c}/K=a/a_{B}$,
where $a_{B}$ is the Bohr radius$(=\hbar^{2}\epsilon/m^{*}e^{2})$,
crystallization is expected for $r_{s}\geq 37$ from Monte Carlo
simulation results \cite{tanatar}.
 Wigner crystallization was first experimentally
observed in two-dimensional sheets of electrons trapped on the surface
of liquid helium\cite{grimes}, where the electron gas is almost classical.  The
two-dimensional electron gas (2DEG) in modulation-doped high mobility
GaAs-AlGaAs heterostructures have proven to be excellent candidates
for observing the formation of a WC in the quantum regime,
largely because such high purity samples are available that
the electrons are not necessarily dominated by disorder effect at the
low densities required to obtain crystalline order. However, in the
absence of a magnetic field, the WC has not been observed
in GaAs 2DEG systems.\cite{pudalov}
Application of
a strong perpendicular magnetic field further enhances the possibility
of forming a WC, since this reduces quantum fluctuations (which
tend to melt the crystal at large electron density), by confining the
electron zero-point motion to cyclotron orbits of radius the
magnetic length, $l_{c}=(\hbar c/eB)^{1/2}$.
Once $l_{c}$ is smaller than the mean
inter-electron distance, Coulomb correlations lead the 2DEG first to form an
incompressible liquid phase (fractional QHE ground state)
at certain densities, and
ultimately to crystallize below some critical filling
factor $\nu(=nhc/eB)=\nu_{c}\ll 1$ at low enough temperature.
Recent studies of high mobility heterojunctions in strong magnetic fields
have uncovered a number of intriguing properties that in some ways are
consistent with the presence of some crystalline order
at the lowest available temperatures.  These include rf
data\cite{andrei,paalanen},
transport experiments\cite{goldman},
cyclotron resonance\cite{besson}, and photoluminescence (PL)
experiments\cite{buhmann,clark}. It is the last of these that
we will discuss theoretically in this paper.

Photoluminescence experiments on these systems have been performed in two ways.
One set of experiments\cite{buhmann} uses a low density of
Be dopants that are purposely grown into the
sample approximately 250\AA~away from the 2DEG.  A pulse of light excites
a core electron out of a Be$^-$ acceptor, and the photoluminescence spectrum
from
recombination of electrons in the 2DEG with the remaining
core hole is observed.  More recent experiments\cite{clark}
have also investigated
recombination of electrons
with itinerant holes in the host crystal (GaAs) valence band.
Both experiments show intriguing and complicated results; among them is
the observation of a pair of photoluminescence lines that appear at magnetic
fields for which transport anomalies recently associated with the WC
are found.  At the lowest temperatures, the lower of the two lines has
most of the oscillator strength; as the temperature is raised, the
oscillator strength transfers to the higher of these lines, until the
lower line cannot be distinguished from the background.  While it is
tempting to associate the lower line with a crystal phase, and the upper with
a melted phase,  the precise interpretation of the data is hampered by
a lack of theoretical understanding of what the PL spectrum should
look like when the ground state of the 2DEG really is a WC.

To address this problem, we have developed a theory of photoluminescence
using a time-dependent Hartree-Fock approximation(TDHFA)\cite{c6} that is
appropriate for the two-dimensional Wigner crystal
in a strong magnetic field.  The cases of localized
and itinerant holes are both studied.
We will show that, within the Hartree-Fock approximation (HFA),
the photoluminescence
spectrum is a weighted measure of the single particle density of states
of the electron system, which for an undisturbed electron lattice has
a very
interesting structure:
for rational filling fractions $\nu=p/q$ there are $q$ subbands,
and in Hartree-Fock, $p$ of these
are filled.  (That the density of states breaks up into $q$ bands
in the simultaneous presence of a periodic potential and magnetic field was
pointed out by Hofstadter\cite{hofstadter},
 and this is often referred to as the
Hofstadter spectrum.)
We therefore expect that for any filling fraction $p/q$, one should
expect to see $p$ lines in the PL for the ideal case of a perfect electron
lattice.  An observation of this behavior in photoluminescence experiments
would yield direct confirmation of the presence of a WC in the system.
Unfortunately, we will find that for the case of a localized hole,
a strong interaction of the hole with the electron lattice tends to wipe out
this structure.
In the case of an unscreened hole (e.g., a valence band
hole in a narrow quantum well),
this arises because the potential localizes an electron
in the vicinity of the hole in the initial state, which dominates
the photoluminescence spectrum.  For the case of a screened hole,
as in a core hole of a neutral acceptor atom,
the initial state of the Wigner crystal is relatively undisturbed.
However, the only final state which is significantly coupled to
via photoluminescence is one in which a vacancy is bound to the
charged acceptor ion in the final configuration.
In both cases, a single final state is strongly favored in the
recombination process, producing a single line in the spectrum.
As a function of temperature, we find that this line shifts upward
in energy within a very narrow range of the melting temperature.
For a system of Wigner crystal domains with a distribution of sizes and melting
temperatures, this would appear as a shifting of oscillator strength
from a low energy photoluminescence line to a high energy one over
a range of temperatures.
Such behavior is in qualitative agreement with experiment\cite{clark}.

For the case of an itinerant hole, which could be generated in a wide
quantum well system,
we find that electron-hole interactions
do not significantly alter the density of states of the Wigner crystal.
One thus sees, in principle, $p$ lines in the
photoluminescence spectrum.  Unfortunately, because of the exchange
interaction, the gaps between these lines
at the relevant small filling fractions
may be extremely
small, making experimental observation of this effect
difficult.  We find an interesting possible
way out of this difficulty, by considering itinerant holes at
finite temperature. For the itinerant case,
the hole moves in a periodic potential (of the electron crystal)
and a magnetic field,
generating a density of states with several bands.  Because there
is no exchange interaction between the electrons and the hole,
we find
gaps in the density of states for the latter which are much larger than
those of the former.  Finite temperature allows a significant probability
of occupying some of the higher hole bands in the initial state, leading
to several new lines in the photoluminescence spectrum.
Once again, observation of this effect would constitute direct
confirmation of crystalline order in the 2DEG.  We believe that, with
improved sample quality, itinerant hole PL experiments
should offer the best opportunity to observe this
interesting behavior.

This paper is organized in the following way. In Sec.II we
show how one can use the
TDHFA to compute the photoluminescence spectrum of this system
for both localized and itinerant
holes. We present and discuss our numerical results in Sec.III.
Finally, we summarize our results and make some concluding remarks
in Sec.IV. A brief account of some of these results has
appeared previously\cite{fertig2}.

\vspace{1cm}

\begin{center}
{\bf II. THEORY}
\end{center}

 In this section we first present a
general expression for the Hartree-Fock Hamiltonian of the 2DEG in a
strong magnetic field. We then derive our theory for photoluminescence from
a WC in this system within TDHFA for both localized and itinerant
holes.

\noindent{\em 1. Model Hamiltonian}

It is well known that non-interacting two-dimensional electrons in a
perpendicular magnetic field($\vec{B}=-B\hat{z}$) have
an energy spectrum of discrete Landau levels:
$E_{N}=(N+\frac{1}{2})\hbar\omega_{c}, N=0,1,2,...$, where
$\omega_{c}=eB/m^{*}c$ is the cyclotron resonance frequency. Working
in the Landau gauge($\vec{A}=-Bx\hat{y}$), and
with periodic boundary conditions in the
$\hat{y}$-direction, the
single particle eigenstates are given by:
\begin{equation}
  <\vec{r}|NX>=\frac{1}{L_{y}}exp(iXy/l_{c}^{2})\phi_{N}(x-X).
\end{equation}
Here $l_{c}=(\hbar c/eB)^{1/2}$ is the magnetic length and $\phi_{N}$
is the one-dimensional harmonic-oscillator eigenstate with oscillation
centers $X$.  The allowed values of $X$ are separated by
$2\pi l_{c}^{2}/L_{y}$. The
degeneracy of each Landau level is given by $g=S/2\pi l_{c}^{2}$, with
$S$ the area of the 2DEG.

In order to derive the theory of photoluminescence from a strong
magnetic field induced two-dimensional WC,
we now consider a system with an interacting 2DEG (within TDHFA) and
a layer with a low density of holes
(either localized or itinerant)
separated by distance $d$, in the presence of a strong perpendicular
magnetic field. Since the magnetic field is very strong; {\em
i.e.,} $\hbar\omega_{c}>>e^{2}/\epsilon a$, and the electronic filling
factor $\nu =nhc/eB<<1$ in the WC regime, we can assume that only the lowest
Landau level is occupied by electrons. The general many-body Halmiltonian can
be expressed in terms of single-electron eigenstates (only $N=0$
eigenstates are included) as follows:
\begin{equation}
 {\cal H}={\cal H}_{o}+{\cal H}_{ee}+{\cal H}_{eh},
\end{equation}
where
\begin{eqnarray}
\label{h123}
 {\cal H}_o & = &
\sum_{X}\frac{1}{2}\hbar\omega_{c}a^{\dag}_{X}a_{X}
+\sum_{i}E_{h}c^{\dag}_{i}c_{i} \nonumber \\
 {\cal H}_{ee} & = & \frac{1}{2S}\sum_{\vec{q}\neq 0}
\sum_{X_{1}X_{2}X_{3}X_{4}}V_{c}(\vec{q})<X_{1}|exp(i\vec{q}\cdot\vec{r})
|X_{4}><X_{2}|exp(-i\vec{q}\cdot\vec{r})|X_{3}>
a^{\dag}_{X_{1}}a^{\dag}_{X_{2}}a_{X_{3}}a_{X_{4}}\nonumber \\
{\cal H}_{eh} & = & \frac{1}{S}\sum_{\vec{q}}\sum_{X_{1}X_{2}}
\sum_{ij}V_{eh}(\vec{q})<X_{1}|exp(i\vec{q}\cdot\vec{r})|X_{2}>
<i|exp(-i\vec{q}\cdot\vec{r})|j>a^{\dag}_{X_{1}}a_{X_{2}}c^{\dag}_{i}c_{j}.
\end{eqnarray}
Here ${\cal H}_{o}$ is the single-particle zero-point energy which is
constant. ${\cal H}_{ee}$ is the electron-electron interaction, in
which $V_{c}(\vec{q})=2\pi e^{2}/\epsilon q$ is the
two-dimensional Fourier transform
of the Coulomb interaction.
${\cal H}_{eh}$ is the electron-hole interaction, in
which $V_{eh}(\vec{q})$ is the Fourier transform of the interaction
between one electron and one hole (for both screened and unscreened
hole). In the case of an itinerant hole, the interaction between electron
and hole is assumed to be weak enough so that it
can be ignored when we calculate the
{\it electronic} density of states.
For a system with a small number of holes, it is a good
approximation to also ignore hole-hole interactions,
which are extremely small in comparison with ${\cal H}_{eh}$.
In our current consideration, we also do not include
the effect of impurities and any external potential.
In practice, we may also drop ${\cal
H}_{o}$ because it is just a constant.
The matrix elements in Eq.(\ref{h123}) are given by
\begin{equation}
<X_{1}|exp(i\vec{q}\cdot\vec{r})|X_{2}>=exp(\frac{i}{2}q_{x}(X_{1}+X_{2})-\frac{
q^{2}l_{c}^{2}}{4})\delta_{X_{1},X_{2}+q_{y}l_{c}^{2}}
\label{xqx}
\end{equation}

 Shown in Fig.1 is the electron density profile of a two-dimensional Wigner
crystal in a strong magnetic field. Here we assume that the two-dimensional
WC is a triangular lattice (which classically is the lowest energy
crystal structure at zero temperature\cite{bonsall}).
Fig.1(a) illustrates a perfect
electronic WC, and Fig.1(b) is for a WC with one localized hole,
at the origin, separated from the plane of the electron
gas by a distance $d(\simeq 250\AA )$. We can
see that the WC deforms around the hole due to the electron-hole
interaction.  (We will explain below how these figures are derived.)
In order to take into account this effect, while still taking advantage of
the periodicity of the WC, we divide the WC system into hexagon unit
cells. Each supercell contains a finite number of electrons, and one
hole (in the center, for localized hole case).
The finite size of our unit cells will not be significant for
large enough supercells.

The density plots illustrated in Fig. 1 were derived using the TDHFA.
We briefly review the salient points of the procedure; details may
be found in Ref.\cite{cote}
We first define the density operator:
\begin{equation}
n(\vec{G})=\int d^{2}\vec{r}exp(-i\vec{G}\cdot\vec{r})n(\vec{r})
=g\rho(\vec{G})exp(-\frac{G^{2}l_{c}^{2}}{4}),
\end{equation}
where
\begin{equation}
\rho(\vec{G})=\frac{1}{g}\sum_{X_{1}X_{2}}e^{-\frac{i}{2}
G_{x}(X_{1}+X_{2})}\delta_{X_{1},X_{2}-G_{y}l_{c}^{2}}a^{\dag}_{X_{1}}
a_{X_{2}}.
\end{equation}
One may easily show that
\begin{equation}
        <\rho(\vec{G}=0)>=\frac{<N_{e}>}{g}=\nu.
\end{equation}
Here $N_{e}$ is the electron number operator.

The Hartree-Fock Hamiltonian for the 2DEG (with an electron-hole
interaction appropriate for a localized hole) in the
lowest Landau level can be
written as:
\begin{equation}
{\cal H}=g\sum{\vec{G}}\left[
W(\vec{G})<\rho(\vec{G})>+
n_{h}V_{eh}(\vec{G})
e^{-G^{2}l_{c}^{2}/4}
\right]\rho(\vec{G})
\end{equation}
where $n_{h}$ is the density of holes.
 $W(\vec{G})$ is the effective Hartree-Fock interaction:
\begin{equation}
W(\vec{G})=\frac{e^{2}}{\epsilon l_{c}^{2}}\left[
\frac{1}{Gl_{c}}e^{-G^{2}l_{c}^{2}/2}(1-\delta_{\vec{G},0})-\sqrt\frac{\pi}{2}
e^{-G^{2}l_{c}^{2}/4}I_{0}(\frac{-G^{2}l_{c}^{2}}{4})\right]
\end{equation}
where $I_{0}(x)$ is the modified Bessel function of the first kind.

\noindent{\em 2. Green's Function, Electron Density, and Density of States:}

We define single electron Green's function:
\begin{equation}
G(X_{1},X_{2};\tau)=-<T_{\tau}a_{X_{1}}(\tau)a^{\dag}_{X_{2}}(0)>
\end{equation}
It is convenient to define the Fourier transform
\begin{equation}
G(\vec{G},\tau)=\frac{1}{g}\sum_{X_{1}X_{2}}e^{-\frac{i}{2}
G_{x}(X_{1}+X_{2})}\delta_{X_{1},X_{2}+G_{y}l_{c}^{2}}
G(X_{1},X_{2};\tau).
\end{equation}
We will use this form of the Fourier transform throughout
this work.
The TDHFA is derived by writing
the equation of motion for the Green's function,
\begin{eqnarray}
\frac{\partial}{\partial\tau}G(X_{1},X_{2};\tau)
&=&-\frac{\partial}{\partial\tau}<T_{\tau}a_{X_{1}}
(\tau)a^{\dag}_{X_{2}}(0)>\nonumber \\
 &= &-\delta_{N_{1}N_{2}}\delta_{X_{1}X_{2}}\delta({\tau})-
 <T_{\tau}[{\cal H}_{eff}-\mu_{e}N_{e},a_{X_{1}}](\tau)a^{\dag}_{X_{2}}(0)>.
\end{eqnarray}
The commutators may be computed explicitly, and
the result is simplified using a Hartree-Fock decomposition.
After Fourier transforming with respect to time,
the equation of motion for
$G(\vec{G},\omega_{n})$ can be written as
\begin{equation}
(i\omega_{n}+\mu_{e})G(\vec{G},i\omega_{n})
-\sum_{\vec{G}'}B(\vec{G},\vec{G}')G(\vec{G}',i\omega_{n})
=\delta_{\vec{G},0},
\label{green}
\end{equation}
where
\begin{equation}
B(\vec{G}_{1},\vec{G}_{2})=\left[
W(\vec{G}_{1}-\vec{G}_{2})\rho(\vec{G}_{1}-\vec{G}_{2})+
n_{h}V_{eh}(\vec{G}_{1}-\vec{G}_{2})
e^{-(\vec{G}_{1}-\vec{G}_{2})^{2}l_{c}^{2}/4}
\right] e^{i(\vec{G}_{1}\times\vec{G}_{2})l_{c}^{2}/2}.
\label{mB}
\end{equation}
We can directly diagonalize matrix $B$ and obtain its eigenvectors
$V_{j}(\vec{G})$ and eigenvalues $\omega^{e}_{j}$,
after which the Green's
function can be written as
\begin{eqnarray}
G(\vec{G},i\omega_{n})&=&\sum_{j}\frac{V_{j}(\vec{G})V_{j}^{*}(\vec{G}=0)}
{i\omega_{n}+\mu_{e}-\omega^{e}_{j}}
\nonumber \\
 & =& \sum_j\frac{W_e(\vec{G},j)}{i\omega_{n}+\mu_{e}-\omega^{e}_{j}}.
\label{green2}
\end{eqnarray}
The density of states for electrons is then given by:
\begin{eqnarray}
D(E)&=&-\frac{1}{\pi}Im(G(\vec{G}=0,E+i\delta))(\frac{g}{S})
\nonumber \\
 &= &
-\frac{1}{\pi}Im(\sum_{j}\frac{W_e(\vec{G}=0,j)}
{E-\omega^{e}_{j}+i\delta})(\frac{1}{2\pi l_{c}^{2}})
\label{dos}
\end{eqnarray}
Finally, the density operator can be expressed as
\begin{eqnarray}
\rho(\vec{G})
 & = & G(\vec{G},\tau =0^{-}) \nonumber \\
 & = &
\sum_{j}V_{j}(\vec{G})V_{j}^{*}(\vec{G}=0)f_{FD}(\omega^{e}_{j}-\mu_{e}).
\label{r17}
\end{eqnarray}
Here $f_{FD}(x)=[1+exp(\beta x)]^{-1}$ is the Fermi-Dirac distribution.
Since $\rho(\vec{G}=0)=\nu $, we can self-consistently calculate
chemical potential $\mu_{e}$, the density of states and the electron
density in $\vec{G}$ space.  The density plots in Fig.1 were obtained
by iteratively solving Eqs.(\ref{green}),(\ref{green2}), and (\ref{r17}).

\noindent{\em 3. Photoluminescence theory for localized hole:}

We now present in detail our theory for photoluminesence from the WC in a
strong magnetic field.
The photoluminescence intensity is given, for a single localized
hole state, by
\begin{equation}
P(\omega)={{I_0} \over {Z}} \sum_n \sum_m e^{-E_n/k_BT}
|<m,0|\hat{L}|n,h>|^2 \delta(\omega-E_n+E_m),
\label{p1}
\end{equation}
where $Z=\Sigma_n e^{-E_n/k_BT}$, $|n,h>$
is a many-body
electron state with energy $E_n$
and $N$ electrons when there is a core hole present,
$|m,0>$ is a many-body electron state with $N-1$ electrons and
energy $E_m$, ~$\omega$ is the luminescence frequency, and
$\hat{L} = \int d^2\vec{r} \psi(\vec{r}) \psi_h(\vec{r})$ is
the luminescence operator,
with $\psi(x)$ the electron annihilation operator and $\psi_h(x)$ the
hole annihilation operator.  As written, the initial state is actually
higher in energy than the final state, and we find it convenient to rework
the problem in terms of absorption rather than emission.  To accomplish
this, we add a term $H^{\prime} = -E_0c_0^{\dag}c_0$ to the Hamiltonian,
where $c_0^{\dag}$ creates a localized hole,
and take the limit $E_0 \rightarrow \infty$.  It is not difficult to show
\begin{equation}
 P(\omega) = \lim_{E_0 \rightarrow \infty}
\frac{P^{\prime}(\omega-E_0)}{n_0(E_0)}
\label{PL1}
\end{equation}
where $P^{\prime}$ is the absorption spectrum of the new Hamiltonian, and
$n_0$ is the average occupation of the hole state, which just becomes one
in the limit $E_0 \rightarrow \infty$.  The absorption spectrum is
identical to Eq.(\ref{p1}), except one needs to add the energy $E_0$ to all the
quantities $E_n$ in the expression.  After standard manipulations\cite{mahan},
one can show that
\begin{equation}
P^{\prime}(\omega) = {{I_0} \over {\pi}} {1 \over {1-e^{\omega/k_BT}}}
Im~R(\omega+i\delta)
\label{PL2}
\end{equation}
The function $R(\omega+i\delta)$ is a response function, which continued
to imaginary frequency has the form
\begin{equation}
 {\cal R}({\it i}\omega_n) = -\int_0^{\beta} <T_{\tau}L(\tau)L^{\dag}(0)>
e^{i\omega_n \tau} d\tau
\label{R1}
\end{equation}

To compute this quantity, we consider (for the case
of a localized hole state) instead of a single hole, a
periodic (hexagonal) lattice of them, with a unit cell that contains as many
electrons as can be handled
numerically.  We allow neither interactions between the holes
nor tunneling between the hole sites, so that in the limit of large
unit (super)cells, one should expect the result to be the same as for the
isolated hole case.

The quantity of interest arising out of Eq.(\ref{R1}) is
\begin{equation}
{\cal R}_{ij} (X_1 ,X_2 ;\tau)=-<T_\tau
a_{X_1}(\tau)c_i(\tau)c_j^{\dag}(0)a_{X_2}^{\dag}(0)>.
\end{equation}
We write down the equation of motion for ${\cal R}_{ij}(X_1,X_2;\tau)$
in terms of its commutator with the Hamiltonian:
\begin{equation}
\begin{array}{cl}
 &\frac{\partial}{\partial\tau}{\cal R}_{ij}(X_{1},X_{2};\tau)
\equiv -\frac{\partial}{\partial\tau}<T_{\tau}a_{X_{1}}
(\tau)c_i(\tau)c_j^{\dag}(0)a^{\dag}_{X_{2}}(0)>\nonumber \\
=& -<[a_{X_1}c_i, c_j^{\dag}a^{\dag}_{X_2}]>\delta(\tau) -<T_\tau
[{\cal H}_{eff}-\mu(N_e-N_h), a_{X_1}c_i](\tau)c_j^{\dag}a_{X_2}^{\dag}>,
\end{array}
\label{Req1}
\end{equation}
where $N_e$ and $N_h$ are corresponding electron and hole number operator.
Here
\begin{eqnarray}
{\cal H}_T & = & {\cal H}_{eff}-\mu(N_e-N_h) \nonumber \\
& = &
{\cal H}_o+{\cal H}_{ee}+{\cal H}_{eh}-E_0\sum_ic_ic_i^{\dag}-\mu(N_e-N_h)
\nonumber \\
 & = & H_1+H_2+H_3,
\end{eqnarray}
where
\begin{eqnarray}
 H_1&=&(\frac{1}{2}\hbar\omega_c-\mu)\sum_Xa_X^{\dag}a_X
+(E_h-E_0+\mu)\sum_ic_i^{\dag}c_i \nonumber \\
 H_2&=&{\cal H}_{ee} \nonumber \\
 H_3&=&{\cal H}_{eh}.
\end{eqnarray}
We can thus write Eq.(\ref{Req1}) as
\begin{equation}
\frac{\partial}{\partial\tau}{\cal R}_{ij}(X_{1},X_{2};\tau)
=-T_0-T_1-T_2-T_3,
\end{equation}
where $T_0$ is the first term on the right hand side of  Eq.(\ref{Req1}),
and the $T_i$'s represent the commutator terms with each
of the $H_i$'s.

In the limit $E_0\rightarrow\infty$, the first term can be written as
\begin{eqnarray}
T_0 &=& [<a_{X_1}a_{X_2}^{\dag}>\delta_{ij}
-<c_j^{\dag}c_i>\delta_{X_1X_2}]\delta(\tau) \nonumber \\
 &=&\lim_{n_0\rightarrow 1} [<a_{X_1}a_{X_2}^{\dag}>
-n_0\delta_{X_1X_2}]\delta_{ij}\delta(\tau) \nonumber \\
 &=&-<a_{X_2}^{\dag}a_{X_1}>\delta_{ij}\delta(\tau)
\end{eqnarray}
Its Fourier transformation is
\begin{eqnarray}
\widetilde{T}_0(\vec{G},\tau)&=&\frac{1}{g}\sum_{X_{1}X_{2}}e^{-\frac{i}{2}
G_{x}(X_{1}+X_{2})}\delta_{X_{2},X_{1}-G_{y}l_{c}^{2}}T_0 \nonumber \\
 &=&-\rho(\vec{G})\delta_{ij}\delta(\tau)
\end{eqnarray}
Since $H_1$ essentially provides a constant
energy shift to the system, the second term can be easily calculated:
\begin{eqnarray}
T_1&=&< T_\tau
[H_1, a_{X_1}c_i](\tau)c_j^{\dag}a_{X_2}^{\dag}> \nonumber \\
 &=& (\frac{1}{2}\hbar\omega_c+E_h-E_0){\cal R}_{ij}(X_{1},X_{2};\tau)
\end{eqnarray}
so that
\begin{equation}
\widetilde{T}_1(\vec{G},\tau)=(\frac{1}{2}\hbar\omega_c+E_h-E_0){\cal
R}_{ij}(\vec{G},\tau)
\end{equation}
In order to calculate the third term (contribution of electron-electron
interaction), we first compute the commutator, and then
apply a Hartree-Fock decomposition technique\cite{cote} to get
\begin{eqnarray}
T_2(X,X^{\prime};\tau)&=&< T_\tau
[H_2, a_Xc_i](\tau)c_j^{\dag}a_{X^{\prime}}^{\dag}> \nonumber \\
 &=& \frac{1}{S}\sum_{\vec{q}\neq 0}
\sum_{X_{1}X_{2}X_{3}X_{4}}V_{c}(\vec{q})<X_{1}|exp(i\vec{q}\cdot\vec{r})
|X_{4}><X_{2}|exp(-i\vec{q}\cdot\vec{r})|X_{3}> \\
 & & \mbox{ }\times \left\{ <a^{\dag}_{X_1}a_{X_4}>\delta_{XX_2}{\cal
R}_{ij}(X_3,X^{\prime};\tau)-<a^{\dag}_{X_1}a_{X_3}>\delta_{XX_2}{\cal
R}_{ij}(X_4,X^{\prime};\tau) \right\}. \nonumber
\end{eqnarray}
Substituting
\begin{equation}
<a^{\dag}_{X_1}a_{X_2}>=
\sum_{\vec{G}}\rho(\vec{G})e^{\frac{i}{2}
G_{x}(X_1+X_2)}\delta_{X_1,X_2-G_{y}l_{c}^{2}},
\label{ara}
\end{equation}
and Eq.(\ref{xqx}), after some lengthy algebra, we obtain the final
expression:
\begin{equation}
T_2(X,X^{\prime};\tau)=\sum_{\vec{G}}W(\vec{G})\rho(\vec{G})e^{iG_x(X-G_yl_c^{2}
/2)}{\cal
R}_{ij}(X-G_yl_c^{2},X^{\prime};\tau)
\end{equation}
Its Fourier transform can be written as:
\begin{equation}
\widetilde{T}_2(\vec{G},\tau)=\sum_{\vec{G}^{\prime}}W(\vec{G}^{\prime})
\rho(\vec{G}^{\prime})e^{-i(\vec{G}\times\vec{G}^{\prime})l_c^{2}/2}{\cal
R}_{ij}(\vec{G}-\vec{G}^{\prime},\tau)
\end{equation}
Now we turn to the fourth term(contribution from electron-hole
interaction). We apply the same Hartree-Fock decomposition technique
and obtain:
\begin{eqnarray}
 \label{t3}
T_3&=&< T_\tau
[H_3, a_Xc_i](\tau)c_j^{\dag}a_{X^{\prime}}^{\dag}> \nonumber \\
 &=& \frac{1}{S}\sum_{\vec{q}}
\sum_{X_{1}X_{2}}V_{eh}(\vec{q})<X_{1}|exp(i\vec{q}\cdot\vec{r})|X_{2}>
\left\{ <a^{\dag}_{X_1}a_{X_2}>e^{-i\vec{q}\cdot\vec{R}_i}{\cal
R}_{ij}(X,X^{\prime}) \right.   \\
& &\mbox{ } \left. +\left[ n_0N_c\sum_{\vec{G}}\delta_{\vec{q},\vec{G}}
\delta_{X_1X}+(1-n_0)\delta_{X_1X}e^{-i\vec{q}\cdot\vec{R}_i}
-<a^{\dag}_{X_1}a_X>e^{-i\vec{q}\cdot\vec{R}_i}\right]
{\cal R}_{ij}(X_2,X^{\prime}) \right\} \nonumber
\end{eqnarray}
Where $N_c$ is the total number of supercells.
In deriving the above equation,
we assume there is no overlap between hole
states ({\em i.e.}, single hole approximation), so
\begin{equation}
<i|exp(-i\vec{q}\cdot\vec{r})|j>=exp(-i\vec{q}\cdot\vec{R}_i)\delta_{ij}
\end{equation}
where $\vec{R}_i$ is the hole (or supercell) superlattice vector.
We then substitute Eq.(\ref{ara}) and Eq.(\ref{xqx}) into
Eq.(\ref{t3}) and take $n_0=1$, and after a very involved calculation, we get
\begin{eqnarray}
T_3(X,X^{\prime};\tau)
&=&\sum_{\vec{q}}V_{eh}(\vec{q})e^{-iq^{2}l_c^{2}/4-iq_xq_yl_c^{2}/2}
\nonumber\\
 & &\mbox{ }\times\left[
\frac{1}{2\pi l_c^2}\sum_{\vec{G}}\rho(\vec{G})
\delta_{\vec{q},-\vec{G}}e^{iG_xG_yl_c^{2}/2}{\cal
R}_{ij}(X,X^{\prime}) \right. \nonumber \\
 & &\mbox{ } +n_h\sum_{\vec{G}}\delta_{\vec{q},\vec{G}}e^{iq_xX}
{\cal R}_{ij}(X-q_yl_c^{2},X^{\prime}) \\
 & &\mbox{ } \left.
-\frac{1}{S}\sum_{\vec{G}}\rho(\vec{G})e^{-i\vec{q}\cdot\vec{R}_i}
e^{iq_x(X-G_yl_c^{2})+iG_x(X-G_yl_c^{2}/2)}{\cal
R}_{ij}(X-(q_y+G_y)l_c^2,X^{\prime}) \right]\nonumber
\end{eqnarray}
in which the last term can be dropped as we take the limit
$S\rightarrow\infty$. Its Fourier transformation can be written as
\begin{eqnarray}
\widetilde{T}_3(\vec{G},\tau)&=&\frac{1}{2\pi
l_c^2}\sum_{\vec{G}^{\prime}}V_{eh}(-\vec{G}^{\prime})\rho(\vec{G}^{\prime})e^{-
{G^{\prime}}^2l_c^2/4}{\cal
R}_{ij}(\vec{G},\tau) \\
 & &\mbox{
}+n_h\sum_{\vec{G}^{\prime}}V_{eh}(\vec{G}^{\prime})e^{-{G^{\prime}}^2l_c^2/4-i(
\vec{G}\times\vec{G}^{\prime})l_c^{2}/2}{\cal
R}_{ij}(\vec{G}-\vec{G}^{\prime},\tau) \nonumber
\end{eqnarray}
So finally, in the $\vec{G}$ space, equation of motion for ${\cal
R}_{ij}$ is
\begin{equation}
\frac{\partial}{\partial\tau}{\cal
R}_{ij}(\vec{G},\tau)=-\widetilde{T}_0-\widetilde{T}_1-\widetilde{T}_2
-\widetilde{T}_3
\end{equation}
When transformed into real frequency space, it becomes
\begin{eqnarray}
-(\omega +i\delta)R_{ij}(\vec{G},\omega) &=&
\rho(\vec{G})\delta_{ij}+(E_0-\frac{1}{2}\hbar\omega_c-E_h)
R_{ij}(\vec{G},\omega) \nonumber \\
 & &\mbox{ }
-\sum_{\vec{G}^{\prime}}W(\vec{G}^{\prime})\rho(\vec{G}^{\prime})e^{-i(\vec{G}
\times\vec{G}^{\prime})l_c^{2}/2}
R_{ij}(\vec{G}-\vec{G}^{\prime},\omega) \\
 & &\mbox{ } -\frac{1}{2\pi
l_c^2}\sum_{\vec{G}^{\prime}}V_{eh}(-\vec{G}^{\prime})\rho(\vec{G}^{\prime})
e^{-{G^{\prime}}^2l_c^2/4}
R_{ij}(\vec{G},\omega) \nonumber \\
 & &\mbox{ }
-n_h\sum_{\vec{G}^{\prime}}V_{eh}(\vec{G}^{\prime})e^{-{G^{\prime}}^2l_c^2/4-
i(\vec{G}\times\vec{G}^{\prime})l_c^{2}/2}
R_{ij}(\vec{G}-\vec{G}^{\prime},\omega) \nonumber
\end{eqnarray}
It is apparent that the solution to this satisfies
\begin{equation}
R_{ij}(\vec{G},\omega)=R(\vec{G},\omega)\delta_{ij}.
\end{equation}
This result may be expressed in the form
\begin{equation}
\sum_{\vec{G}^{\prime}}\biggl[(\omega+i\delta+E_0-\omega_0)
\delta_{\vec{G},\vec{G}^{\prime}}-B(\vec{G},\vec{G}^{\prime})\biggr]
R(\vec{G}^{\prime},\omega) = -\rho(\vec{G}),
\label{emR}
\end{equation}
where
\begin{equation}
\omega_0 = \frac{1}{2}\hbar\omega_c+E_h+{1 \over {2\pi\l_c^2}}\sum_{\vec{G}}
\rho(\vec{G})V_{eh}(-\vec{G})e^{-G^2l_c^2/4},
\end{equation}
and $B$ is exactly given by Eq.(\ref{mB}).
Finally, it is not difficult to show that
\begin{equation}
 R(\omega) =
{{n_h S} \over {2\pi l_c^2}}\sum_{\vec{G}} R({\vec{G}},\omega)
e^{-G^2l_c^2/4},
\end{equation}
and the PL spectrum may now be computed using Eqs.(\ref{PL1}) and (\ref{PL2}).

We see that the form of $R$ is essentially that of electron Green's
function as in Eq.(\ref{green}).
By inverting Eq.(\ref{emR}), we have
\begin{eqnarray}
R(\vec{G},\omega)&=&-\sum_{\vec{G}^{\prime}}\biggl[(\omega+i\delta+E_0-\omega_0)
\delta_{\vec{G},\vec{G}^{\prime}}-B(\vec{G},\vec{G}^{\prime})\biggr]^{-1}
\rho(\vec{G}^{\prime})\nonumber \\
 &=&-\sum_{j\vec{G}^{\prime}}\frac{V_j(\vec{G})V_j^{-1}(\vec{G}^{\prime})
\rho(\vec{G}^{\prime})}{\omega+i\delta-\omega_0-\omega_j^e} \\
&=&-\sum_j\frac{W_e(\vec{G},j)f_{FD}(\omega_j^e-\mu_e)}{\omega+i\delta-\omega_0-
\omega_j^e}.
\end{eqnarray}
Here we have already dropped $E_0$ because it cancells out when we calculate
the final photoluminescence power using Eqs.(\ref{PL1}) and (\ref{PL2}).
We can see that $R$ has poles at precisely
the same energies as the poles
 in the electron Green's function for the system in the
presence of the external interaction $V_{eh}$ due to the hole\cite{fertig},
up to the constant energy shift $\omega_0$\cite{c7,c8}.
Thus, the photoluminescence spectrum is indeed a
weighted measure of the single particle  density of states of the
electron system with a localized hole.

\noindent{\em 4. Photoluminescence theory for itinerant hole:}

The case of the itinerant hole is treated similarly to the case outlined
above, except that
there is an important simplification: since the hole density
is low at all points in space, it is safe to ignore any deformation of
the electron lattice due to the hole.
Since the itinerant hole is moving (in a layer a
distance $d$ away from the electron layer) in the periodic potential
of the electron lattice, there will be many hole states forming
several bands similar to the density of states of the electronic
Wigner crystal.
For simplicity, we work in the lowest Landau level of the
hole, although in real systems the relatively heavy hole
mass would allow some (possibly significant) Landau level mixing.
Ignoring this, however, should not affect our qualitative conclusions.
We can define single particle Green's function for the itinerant
hole:
\begin{equation}
G_h(X_1,X_2)=-<T_{\tau}c_{X_1}(\tau)c^{\dag}_{X_2}(0)>.
\end{equation}
We can also use TDHFA to derive the equation of motion for itinerant
hole Green's function. In the momentum space,
$G_h^*(\vec{G},\omega_n)$ satisfies
\begin{equation}
(i\omega_{n}+\mu_{h})G_{h}^{*}(\vec{G},\omega_{n})-\sum_{\vec{G'}}B_{h}(\vec{G},
\vec{G'})G_{h}^{*}(\vec{G'},\omega_{n})=\delta_{\vec{G},0},
\end{equation}
where
\begin{equation}
B_{h}(\vec{G},\vec{G'})=-\frac{e^{2}}{\epsilon
l_{c}}\frac{1}{Gl_{c}}(1-\delta_{\vec{G},0})e^{-Gd-G^{2}l_{c}^{2}/2+i(\vec{G}
\times\vec{G'})l_{c}^{2}/2}\rho(\vec{G}).
\end{equation}
We solve the eigenvalue problem:
\begin{equation}
\sum_{\vec{G'}}B_{h}(\vec{G},\vec{G'})V_{h}(\vec{G'},j)=\omega_{j}^{h}V_{h}
(\vec{G},j),
\end{equation}
so that
\begin{eqnarray}
G_{h}(\vec{G},\omega_{n})&=&\sum_{j}\frac{V_{h}^{*}(\vec{G},j)V_{h}(\vec{G}=0,j)
}{i\omega_{n}+\mu_{h}-\omega_{j}^{h}}
\nonumber \\
 &=&
\sum_{j}\frac{W_{h}(\vec{G},j)}{i\omega_{n}+\mu_{h}-\omega_{j}^{h}},
\end{eqnarray}
and density of states(DOS) for the itinerant holes is
\begin{eqnarray}
D_{h}(E)&=&-\frac{1}{\pi}ImG_{h}(\vec{G}=0)(\frac{g}{S}) \nonumber \\

&=&-\frac{1}{\pi}Im\sum_{j}\frac{W_{h}(\vec{G}=0,j)}{E+i\delta-\omega_{j}^{h}}(
\frac{1}{2\pi l_c^2}).
\end{eqnarray}
The number of bands for the hole density of states should be the same
as that for electron but the band gaps should be quantitatively
different, since the hole-hole
interactions do not play a significant role in the low hole density.

Consider first a single itinerant hole at energy level
$\omega^h_j$, from which we
calculate the photoluminescence power.
Since the hole and the electron have opposite charge (but different effective
mass), their lowest Landau level wavefunction are just complex
conjugates of each other; {\it i.e.,} $\phi^h_X(\vec{r})=\phi_X^*(\vec{r})$.
So the luminescence operator can be written as
\begin{equation}
\hat{L}=\int d^2\vec{r}\psi(\vec{r})\psi_h(\vec{r})=\sum_{X_1X_2}\int
d^2\vec{r}\phi_{X_1}(\vec{r})\phi_{X_2}^*(\vec{r})a_{X_1}c_{X_2}
\end{equation}
Then the electron-hole recombination process is given approximately by
\begin{equation}
 -<T_{\tau}\hat{L}(\tau)\hat{L}^{\dag}(0)>
 =-\sum_{X_1X_2}G(X_1,X_2;\tau)G_h(X_1,X_2;\tau).
\end{equation}
Here,
the single hole Green's
function may be written as
\begin{equation}
G_h(\vec{G},\tau)=W_h(\vec{G},j)e^{-(\omega^h_j-E_0)\tau},
\end{equation}
so that the photoluminescense power for single hole at energy level
$\omega^h_j$ can be written as
\begin{equation}
P_j(\omega)\propto -I_0\sum_{\vec{G}} Im\left[ G(\vec{G},\omega
-\omega^h_j)W_h(\vec{G},j)\right].
\end{equation}
Since there are many hole states
close in energy on the scale of temperature for the itinerant hole,
we need to  take a thermal average over the different hole states
that the electrons may decay into. So the final photoluminescence
power in itenerant hole case is given by:
\begin{equation}
P(E) =
-I_0\sum_{\vec{G}}\sum_{ij}Im\frac{W_e(\vec{G},i)
f_{FD}(\omega_i^e-\mu_e)}{E+i\delta
-\omega_0
-\omega_i^e-\omega_j^h}\mbox{ }
\frac{W_h(\vec{G},j)e^{-\beta\omega_j^h}}{\sum_jW_h(\vec{G}=0,j)e^{-\beta\omega_
j^h}}
\end{equation}
Here $\omega_0=\frac{1}{2}\omega_c+\frac{1}{2}\omega_c^h+\Delta$,
where $\omega_c^h$ is hole cyclotron frequency which is much
smaller than that for electron due to the heavier hole effective mass, and
$\Delta $ is the conduction-band-valence-band gap.

\vspace{1cm}

\begin{center}
{\bf III. NUMERICAL RESULTS}
\end{center}

 From the theoretical analysis in previous section, we observe that
the photoluminescence power just depends on the ground-state electron
density $\rho(\vec{G})$, which we can iteratively calculate by directly
diagonalizing the matrix $B$. We can then determine the electron density of
states(DOS) and calculate the photoluminescence power for the localized
hole case. For the itinerant hole situation, we can also calculate the
hole DOS by
directly diagonalizing the matrix $B_h$, and then calculate the thermal
averaged photoluminescence power. In this section, we present our
numerical results for the electron ground state density profile,
electron density of states (including DOS for itinerant hole), and the
photoluminescence spectra at different temperatures for a
magnetic field-induced Wigner crystal.

As shown in sec.II, the photoluminescence spectrum is essentially a
weighted measure of the density of states of the electron system.
As noted earlier, the DOS for a perfect WC is a
Hofstadter butterfly.
For fractional filling factor $\nu =p/q$, the DOS has $q$ subbands,
and at zero temperature,
the lowest $p$ bands are filled.
We should expect to observe this structure; {\it i.e.}, $p$
 lines for $\nu =p/q$, in the photoluminescence spectrum
of an ideal Wigner crystal if we turn off the electron-hole interaction.
Presented in Fig.2 are
the density of states (a) and the photoluminescence spectrum (b) for an
undisturbed Wigner crystal at $\nu =2/7$\cite{c2}
 with no e-h interaction. We can
clearly see that the DOS has 7 bands, with only the lowest 2 occupied at low
temperature. As expected, the photoluminescence has 2 peaks with
a splitting identical to that of the
DOS. An observation of this behavior in experiments
would directly confirm the presence of a Wigner crystal in the system.

Unfortunately, the PL spectrum in Fig.2 is not experimentally observable
for the case of a localized hole, because of the
strong interaction of the hole with the electron lattice.
We present in Fig.3 the results for
the case of an unscreened hole (e.g., a valence band
hole in a narrow quantum well). We can see, as shown in
Fig.3(b), that there is a
single photoluminescence peak which is shifted down in energy from
the perfect Wigner crystal case. This structure is best interpreted in
terms of the density of states as illustrated in Fig.3(a), for a
periodic electron system with 12 electrons and one localized hole
per unit cell. Because of the attractive electron-hole
interaction,
the WC is deformed in the vicinity of the hole, as shown in
Fig.1(b).  The two filled subbands in the density of states breaks up into
three bands, with the lowest energy peak much smaller in weight but shifted
down in energy. The lowest energy peak corresponds to a bound
state of the electrons with the hole, and because this is the only
state with a significant overlap with the hole, it
dominates
the photoluminescence spectrum.  To show this,
we present in Fig.4 the electron density
profile which corresponds to each energy subband in the DOS. As shown
in Fig.4(a), the lowest energy peak contains a single electron,
localized in the center of the unit cell right above the
hole. Since the photoluminescence power is proportional to the overlap
between electron and hole wavefunctions, this localized electron will
certainly overwhelm the PL spectrum.
We find that the other peaks in the
DOS correspond to sets of electron states that are successively
further away from the hole, for increasing energy, as
 shown in Fig.4(b) and (c). The contribution of these states
to the PL is nearly
2 orders of magnitude smaller than that of the localized electron in the
center.

It should be noted that, in most localized hole experiments\cite{buhmann},
the dopant atom is a {\it neutral} acceptor in its initial state.
The interaction of the core hole with the electron gas is then quite weak,
leading to a negligible deformation of the
WC in its initial state, as shown in Fig.5(a).
However, the {\it final} state of the
dopant is charged, which introduces a strong perturbation in the
final state of the WC.
Our Hartree-Fock approach does not handle this situation well, as
it tends to give qualitatively unrealistic energies when there
are strong electron-hole interactions in the final state.
It is easy to see, however, that
the
PL spectrum is still dominated by a single final state, one in
which a vacancy is bound to the charged ion, as shown in Fig.5(b).  The
PL spectrum should be
thus qualitatively the same as described for the case of a
strong initial interaction, {\it i.e.} a single peak will dominate the
PL spectrum. We calculate the PL peak energy simply by finding the
difference between the Hartree-Fock energies of the initial and
final states. The temperature dependence of this PL energy is
presented in Fig.5(c).

Our calculated PL spectrum for a localized hole as the temperature
is raised so as to melt the WC
also has very interesting behavior\cite{comment,c5,c9}.
 As seen in Fig.3 (unscreened hole)
and Fig.5(screened hole), there is an upward shift in the PL peak.
The increase in energy
corresponds directly to the potential energy lost per electron when the
carriers are no longer crystallized.  What is remarkable about the
shift is that it occurs almost precisely at the melting temperature;
there is very little motion just above or below the transition.
This is in qualitative agreement with experimental observations,
in which two distinguishable lines are observed, with oscillator strength
transferring from the lower to the upper one as the temperature is increased.
One could interpret this as finite size domains of the WC with a distribution
of melting temperatures, accounting for the continuous transfer of
oscillator strength between the two lines.  That two such lines are visible
in real experiments, rather than a broad continuum PL spectrum, seems
consistent with an electrostatic environment for
the recombining electrons that
is fairly uniform through the sample, indicating that there may be some
(substantial) order in the system.

The difficulty of observing local crystalline order directly in PL
for the highly localized hole is clearly
related to the fact that a single electron dominates the electron-hole
recombination.  This problem can be alleviated in principle if the hole
is not so strongly localized. We thus
consider an itinerant hole in the valence band, a geometry
which is only very recently being examined in the WC regime\cite{clark}.
As explained above, electron-hole interactions
do not significantly alter the density of states of the Wigner crystal
in this case.
Typical PL spectrum for this system are shown
in Fig. 6(c), for filling fraction $\nu=2/11$. As shown in Fig.6(a),
the electron density of states has 11 bands with 9 bands above the
chemical potential empty, but the splitting between the lowest 2
filled subbands  is too small to resolve numerically\cite{yoshioka}.
This is apparently due to exchange effects.  Thus, for
the lowest temperatures, one only sees a single peak in the
photoluminescence.
However, an interesting effect occurs when the temperature is raised slightly
(although not nearly enough to melt the crystal):  one then finds that
structure is introduced in the PL peak.  This turns out to be due to the
density of states for the hole. The hole also moves in the periodic
potential of the electron lattice and the strong magnetic field,
and so should be expected to
 have eleven bands as well, as seen in Fig.6(b).
Since there is no exchange interaction between the hole and the electrons,
the hole DOS is
qualitatively different from that of the electrons. Specifically,
the gap between the lowest two bands of the hole
is much larger than that of the electrons,
but much smaller than the bandgap between occupied and unoccupied
electron states of the WC.
So if we increase the temperature moderately so as to
allow some non-negligible probability for the hole to occupy the
higher bands, but the electrons remain in the lowest two bands,
each hole subband that has a significant probability of
occupation will add
a new line to the PL spectrum.
Once again, observation of this effect would constitute direct
confirmation of crystalline order in the 2DEG.  We believe that, with
increasing sample quality, this effect should be observable
in wider-well geometries.

\vspace{1cm}

\begin{center}
{\bf IV. CONCLUSION}
\end{center}

In summary, we have developed a theory of photoluminescence for the WC in
a strong magnetic field using the time-dependent Hartree-Fock
approximation. We find that the PL spectrum is a weighted
measure of the DOS of the electron system.
One can use PL to unambiguously
demonstrate the presence of a WC, by observing a gap structure associated
with the unique energy spectrum of an electron in a periodic potential
and a magnetic field.
We show that electron-hole interactions arising from a localized hole
(both screened and unscreened) tend
to remove the Hofstadter structure from the PL spectrum.
Instead, the PL is dominated by a single
peak, arising from a single electron localized
in the vicinity of an unscreened hole. Similarly, for the screened hole,
a vacancy bound to the charged acceptor ion is the only final state
significantly contibuting to the PL, resulting again in a single line.
The behavior of the PL spectrum for the WC in the case of a localized hole
at finite temperature was also
investigated. We found an upward shift in PL energy as we approach
melting point of the WC which qualitatively agrees with the experiment.
We also argued that in an itinerant hole experiment, the gap
structure will be observed by raising the temperature because the hole
itself has a Hofstadter density of states, due to the
periodic potential provided by the
electron WC and the magnetic field.
We believe that, with
improved sample quality, itinerant hole PL experiments
should offer the best opportunity to observe
this type of structure,
which is a direct consequence of the presence of a WC.

Finally, it is important to note that any real sample inevitably has disorder,
so that one should expect these systems to form domains
separated by grain boundaries (provided the disorder is not
strong enough to completely eliminate local crystalline order.)
The effects of this disorder most likely
would be to fill in the gaps in the density of states that are
associated with the Hofstadter butterfly.  Clearly, one needs a very high
quality sample to see direct indications of crystal order in the PL.

\begin{center}
{\bf ACKNOWLEDGMENTS}
\end{center}

The authors thank the University of Kentucky Center for Computational Sciences
for providing computer time.  This work is supported by the National
Science Foundation through grants No.
DMR 92-02255(HAF) and 91-23577(DZL and SDS).

\pagebreak


\pagebreak

\begin{center}
{\bf FIGURE CAPTIONS}
\end{center}

{\bf Fig.1} The density profile for a strong magnetic field induced
Wigner crystal at $\nu =2/7$. (a) For perfect WC with no electron-hole
interaction; (b) WC with 12 electrons and one localized hole per
super-cell in the presence of unscreened electron-hole interaction,
where $b$ is the super-lattice constant.

{\bf Fig.2} Electron density of states and PL spectrum for perfect
WC with no electron-hole interaction for
$\nu =2/7$. (a) Electron DOS below chemical potential
at zero temperature (inset: DOS above chemical potential), where
$E_e=\frac{1}{2}\omega_c$;
(b) PL spectrum at different temperatures:
$T=0.0045T_{melt}$(solid), $T=0.45T_{melt}$(dotted),
$T=0.9T_{melt}$(dashed) and $T=1.12T_{melt}$(inset: dash-dotted).

{\bf Fig.3} Electron density of states and PL spectrum for
WC in the presence of unscreened interaction between electrons and
localized holes for
$\nu =2/7$. (a) Electron DOS below chemical potential
at zero temperature (inset: DOS above chemical potential);
(b) PL spectrum at different temperatures:
$T=0.0045T_{melt}$(solid), $T=0.45T_{melt}$(dotted),
$T=0.9T_{melt}$(dashed) and $T=1.01T_{melt}$(dash-dotted).

{\bf Fig.4} Electron density profile corresponding to different
subbands in DOS (a) the lowest energy peak; (b) the second band; (c)
the third band.

{\bf Fig.5} Results for screened localized hole situation. (a)
Electron density profile for the initial state with weak interaction
between electrons and a core hole in a neutral acceptor atom; (b)
Electron density profile for the final state with a vacancy bound to
charged acceptor ion; (c) Temperature dependence of the PL energy.
Note that dip in the PL just below the melting temperature is a finite
size effect.

{\bf Fig.6} Results for itinerant hole at $\nu =2/11$. (a) Electron
density of states below chemical potential
(inset: DOS above chemical potential), where $E_e=\frac{1}{2}\omega_c$;
(b) Hole density of states, where $E_h=\frac{1}{2}\omega_c^h$;
 (c) PL spectrum at
different temperatures: $T=0.0045T_{melt}$(solid), $T=0.045T_{melt}$(dotted),
$T=0.45T_{melt}$(dashed).


\begin{thebibliography}{99}

\bibitem{wigner} E.P. Wigner, Phys. Rev. {\bf 46}, 1002 (1934)

\bibitem{tanatar} B. Tanatar and D.M. Ceperley, Phys. Rev. B {\bf 39},
5005 (1989)

\bibitem{grimes} C.C. Grimes and G. Adams, Phys. Rev. Lett. {\bf 42},
795 (1979)

\bibitem{pudalov} Some recent experimental work has suggested the
possibility that the WC is observable at zero field in MOSFET's.
See V.M. Pudalov, M. D'Iorio, S.V. Kravchenko, and J.W. Campbell,
Phys. Rev. Lett. {\bf 70}, 1866 (1993)

\bibitem{andrei} E.Y. Andrei, G. Deville, D.C. Glattli, F.I.B. Williams,
E. Paris, and B. Etienne, Phys. Rev. Lett. {\bf 60}, 2765 (1988);
F.I.B. Williams et al., Phys. Rev. Lett.
{\bf 66}, 3285 (1991)

\bibitem{paalanen} M.A. Paalanen, R.L. Willett, P.B. Littlewood, K.W. West,
L.N. Pfeiffer, and D.J. Bishop, Phys. Rev. B {\bf 45}, 11342 (1992)

\bibitem{goldman} V.J. Goldman, M. Santos, M. Shayegan, and J.E. Cunningham,
Phys. Rev. Lett. {\bf 65}, 2189 (1990); H.W. Jiang, R.L. Willett, H.L. Stormer,
D.C. Tsui, L.N. Pfeiffer, and K.W. West, Phys. Rev. Lett. {\bf 65},
633 (1990); Y.P. Li, T. Sajoto, L.W. Engel, D.C. Tsui, and M. Shayegan,
Phys. Rev. Lett. {\bf 67}, 1630 (1991)

\bibitem{besson} M. Besson, E. Gornick, C.M. Engelhardt, and G. Weimann,
Semiconductor Sci. Technol. {\bf 7}, 1274 (1992)

\bibitem{buhmann} H. Buhmann et al., Phys. Rev. Lett. {\bf 66}, 926 (1991);
I.V. Kukushkin {\it et al}, Phys. Rev. B {\bf 45}, 4532 (1992)

\bibitem{clark} E.M. Goldys et al., Phys. Rev. B {\bf 46}, 7957 (1992);
R.G. Clark, Physica Scripta {\bf T39}, 45 (1991) and
references therein.

\bibitem{c6} A mean-field approach cannot account for correlation effects
associated with the Fermi-edge singularity encountered in PL in metals.
However, since the WC has no Fermi surface, this is not a
problem for our present purpose.

\bibitem{hofstadter} D. Hofstadter, Phys. Rev. B {\bf 14}, 2239 (1976)

\bibitem{fertig2} H.A. Fertig, D.Z. Liu, and S. Das Sarma,
Phys. Rev. Lett. {\bf 70}, 1545 (1993)

\bibitem{bonsall} L. Bonsall and A.A. Maradudin, Phys. Rev. B {\bf
15}, 1959 (1977)

\bibitem{cote} R. C\^ot\'e and A.H. MacDonald, Phys. Rev. Lett.
{\bf 65}, 2662 (1990);
Phys. Rev. B {\bf 44}, 8759 (1991)

\bibitem{mahan} G. Mahan, {\it Many-Particle Physics}, (Plenum Press, New York,
1983)

\bibitem{fertig} H.A. Fertig, R.  C\^ot\'e, A.H. MacDonald, and S. Das Sarma,
Phys. Rev. Lett. {\bf 69}, 816 (1992)

\bibitem{c7} However, the residues of the poles are not the same as for
the Green's function.  The residues are determined by the overlaps of
the mean-field single particle electron wavefunctions with the
hole wavefunction.  It is thus appropriate to think of $R$ as a weighted
Green's function.  Since $P^{\prime}(\omega)$ involves the imaginary
part of ${\rm R}(\omega)$, our final result takes the form of a weighted
density of states.

\bibitem{c8} Because we have taken the limit $E_0 \rightarrow \infty$, we
always
consider situations in which the inital state of the hole is occupied
with probability 1; we thus do not consider excitonic correlations between
the electrons and the hole, which would enter here as a non-zero
expectation value $<\psi(0)\psi_h(0)>$, for a hole tightly bound to a
site at the origin.  This should be valid so long as the hole is not too
close to the electron gas on the scale of the magnetic length, as is
the case in most experiments.  See  A.H. MadDonald, E.H. Rezayi, and
D. Keller, Phys. Rev. Lett. {\bf 68}, 1939 (1992)

\bibitem{c2} The filling $\nu=2/7$ is higher than the fillings at which
the WC is believed to be the groundstate of the system.  However, our results
should be qualitatively the same as for lower fillings, and the use of a
larger $\nu$ allows us to obtain better numerical accuracy.  Results for
an experimentally relevant filling ( $\nu=2/11$) for
the itinerant hole case are given in Fig. 6.

\bibitem{comment} Because of the mean-field nature of the Hartree-Fock
approximation, our calculation does not describe the melting transition
when the magnetic field is adjusted so that the electrons melt into
a fractional quantum Hall fluid.  The data for this melting transition
are also quite different than the temperature-driven melting transition,
suggesting that the physics of the two types of transitions is quite
different.  In this work, we address only temperature-driven melting.

\bibitem{c5} The Hartree-Fock approach to melting cannot account for
fluctuation
effects that might lead, for example, to a Kosterlitz-Thouless transition.
However, since the PL for a core hole is a probe of {\it local} electron
structure, we believe our calculation gives the correct qualitative behavior.

\bibitem{c9} Excited phonon modes at finite temperatures strongly renormalize
the electron density modulations, leading to lower melting temperatures
than those calculated in the TDHFA, as well as a broadening of the peak
structures
in all our figures.  However, we do not expect this to affect any of our
qualitative conclusions.

\bibitem{yoshioka} D. Yoshioka and P.A. Lee, Phys. Rev. B {\bf 27},
4986 (1983),
found a similarly small splitting at this filling.

\end{thebibliography}
\end{document}